\documentstyle[11pt,procsla]{article}

\catcode`\@=11
\long\def\@makefntext#1{
\protect\noindent \hbox to 3.2pt {\hskip-.9pt
$^{{\ninerm\@thefnmark}}$\hfil}#1\hfill}		

\def\@makefnmark{\hbox to 0pt{$^{\@thefnmark}$\hss}}  
	
\def\ps@myheadings{\let\@mkboth\@gobbletwo
\def\@oddhead{\hbox{}
\rightmark\hfil\ninerm\thepage}
\def\@oddfoot{}\def\@evenhead{\ninerm\thepage\hfil
\leftmark\hbox{}}\def\@evenfoot{}
\def\sectionmark##1{}\def\subsectionmark##1{}}

\newcommand{\beq}{\begin{equation}}
\newcommand{\eeq}{\end{equation}}

\newcommand{\nn}{\nonumber}
\newcommand{\be}{\begin{equation}}
\newcommand{\ee}{\end{equation}}
\newcommand{\ba}{\begin{eqnarray}}
\newcommand{\ea}{\end{eqnarray}}
\newcommand{\baz}{\begin{eqnarray*}}
\newcommand{\eaz}{\end{eqnarray*}}

\newcommand{\epm}{\varepsilon_{\mu\nu\lambda\sigma}}

\begin{document}

\centerline{\normalsize\bf ON THE TWIST--2 CONTRIBUTIONS TO POLARIZED
}
\baselineskip=22pt
\centerline{\normalsize\bf STRUCTURE FUNCTIONS~\footnote{Talk 
presented by J. Bl\"umlein on the
{\sf International Workshop on Deep Inelastic Scattering}, Rome, April
1996.}}

\centerline{\footnotesize JOHANNES BL\"UMLEIN$^a$
 and  NIKOLAJ KOCHELEV$^{a,b}$
}
\vspace*{0.2cm}
\baselineskip=13pt
\centerline{\footnotesize\it $^a$~DESY--Zeuthen, Platanenallee 6,
D--15735 Zeuthen, Germany}
\vspace*{0.2cm}
\baselineskip=13pt
\centerline{\footnotesize\it $^b$~Bogoliubov Laboratory of Theoretical
Physics,
JINR, RU--141980 Dubna, Moscow Region, Russia}
\centerline{\footnotesize e-mail: blumlein@ifh.de,
 kochelev@thsun1.jinr.dubna.su}

\vspace*{0.9cm}
\abstracts{
The twist--2 contributions to the polarized structure functions
in deep inelastic lepton--hadron scattering are calculated including
the exchange of weak bosons and using both
the operator product expansion and the covariant parton model.
A new relation between two structure functions leading to a
sequence of new sum rules is found.
The light quark mass corrections to the structure functions are derived
in lowest order QCD.
}
\normalsize\baselineskip=15pt
\setcounter{footnote}{0}
\renewcommand{\thefootnote}{\alph{footnote}}
\nopagebreak
\section{Introduction}
\noindent
The study of polarized deep inelastic scattering off polarized targets
has revealed a rich structure of phenomena during the last
years~\cite{REV}. So far only the case of deep inelastic photon
scattering has been studied experimentally.  Future polarized proton
options at  high energy colliders as
RHIC and HERA would allow to probe the spin
structure of nucleons at much higher $Q^2$~(cf.~ref.~\cite{JB95A}) also.
In this
range $Z$--exchange contributions become relevant and one may investigate
charged current scattering as well.
For this general case
the scattering cross section is  determined by (up to) five polarized
structure
functions per current combination.

In
the present paper we derive
the relations for the complete
set of the polarized structure functions including weak interactions
in lowest order QCD
which are not associated with terms in the
scattering cross section
vanishing as $m_{lepton} \rightarrow 0$.
The calculation is performed applying two different
techniques:~the operator product
expansion and  the covariant parton model~\cite{LP}.
In
the latter approach
furthermore
 also
the quark mass corrections are obtained.

As it turns out the twist--2 contributions for
only
two out of the five polarized structure
functions, corresponding to the respective current combinations, are
linearly independent. Therefore three
linear operators have to exist which
determine the remaining three structure
functions over a basis of two in lowest order QCD.
Two of them are given by the
Wandzura--Wilczek\cite{WW} relation  and a
relation by Dicus~\cite{DIC}.
A third {\it new}
relation is found~\cite{BK1}.

We construct
the hadronic tensor
using both Lorentz and time reversal
invariance and current conservation. It is given by:
\be
W_{\mu\nu}^{ab}=\frac{1}{4\pi}\int d^4xe^{iqx}
\langle pS\mid[J_\mu^a(x),J_\nu^b(0)]\mid pS\rangle,
\label{eqHAD}
\ee
where in framework of the quark model the currents are
\be
J_\mu^a(x)=\sum_{f,f'}
U_{ff'} \overline{q}_{f'}(x)\gamma_\mu(g_V^a+g_A^a\gamma_5)q_f(x).
\ee
In terms of structure functions the hadronic tensor reads:
\ba
W_{\mu\nu}^{ab}
&=&(-g_{\mu\nu}+\frac{q_\mu q_\nu}{q^2})F_1^i(x,Q^2)+
\frac{\widehat{p}_\mu\widehat{p}_\nu}{p.q} F_2^i(x,Q^2)-
 i\epm\frac{q_\lambda p_\sigma}{2 p.q}  F_3^i(x,Q^2)\nn\\
&{+}& i\epm\frac{q^\lambda S^\sigma}{p.q} g_1^i(x,Q^2)+
i\epm\frac{q^{\lambda}(p.q S^\sigma - S.q p^\sigma)}
{(p.q)^2} g_2^i(x,Q^2)\nn\\
&{+}& \left[ \frac{\widehat{p_\mu} \widehat{S_\nu}
+ \widehat{S_\mu} \widehat{p_\nu}}{2}-
S.q \frac{\widehat{p_\mu} \widehat{p_\nu}}{(p.q)} \right]
\frac{g_3^i(x,Q^2)}{p.q}\nn\\
&+&
S.q \frac{\widehat{p_\mu}\widehat{p_\nu}}{(p.q)^2}
g_4^i(x,Q^2)+
(-g_{\mu\nu}+\frac{q_\mu q_\nu}{q^2})\frac{(S.q)}{p.q} g_5^i(x,Q^2),
\label{eqz4}
\ea
with  $ab \equiv i$
and
\be
\widehat{p_\mu} = p_\mu-\frac{p.q}{q^2} q_{\mu},~~~~~~\widehat{S_\mu}
= S_\mu-\frac{S.q}{q^2} q_{\mu}.
\label{eqz5}
\ee
Here $x = Q^2/2p.q \equiv Q^2/2M\nu$ and $Q^2 = -q^2$ is the transfered
four
momentum  squared.
$p$ and $S$ denote the four vectors of the
nucleon momentum and spin, respectively, with
$ S^2=-M^2$ and, $S.p = 0 $.
$g_{V_i}$ and $g_{A_i}$ are the vector and axialvector
couplings of the bosons exchanged in the respective subprocesses.
Different other
assignements of structure
functions were used by other authors~(cf.~\cite{BK1,BK2} for a
corresponding survey). They can be obtained as linear combinations of
those defined in eq.~(\ref{eqz4}).

%
\section{Operator Product Expansion}
%
\noindent
We consider the spin--dependent part of the
forward Compton amplitude, $T_{\mu\nu}^{ij, spin}$. It is
related to the corresponding part of the
hadronic tensor by
\be
W_{\mu\nu}^{ij, spin} = \frac{1}{2\pi} Im T^{ij, spin}_{\mu\nu},
\label{eqB1}
\ee
where
\be
T_{\mu\nu}^{ij, spin}
= \left.
i\int d^4xe^{iqx}\langle pS\mid(T{J_\mu^i}^\dagger (x)J_\nu^j(0))\mid
pS\rangle\right |_{spin}.
\label{eqB2}
\ee
The forward Compton amplitude is expanded into local operators near
the light cone and the operator expectation values are calculated.
They are related to the moments of the structure functions introduced
in eq.~(\ref{eqz4}).
In the present paper we are considering only the operators of twist 2.
A complete account of the contributions of the operators of both twist~2
and 3 is given in ref.~\cite{BK2}.
The following relations are obtained:
\ba
\int_0^1 dx x^n g_1^j(x,Q^2) &=& \frac{1}{4}
\sum_q \alpha_j^q a_n^q,~~~{\ } n=0,2...,
\label{eqYMP}
\\
\int_0^1 dx x^n g_2^j(x,Q^2) &=&  - \frac{1}{4}
\sum_q \alpha_j^q
      \frac{n a_n^q}{n + 1} ,~~~{\ } n=2,4...,   \\
\int_0^1 dx x^n g_3^j(x,Q^2) &=&
 \sum_q  \beta_j^q
      \frac{a_{n+1}^q}{n + 2} ,~~~{\ } n=0,2...,   \\
\int_0^1 dx x^n g_4^j(x,Q^2) &=&
\frac{1}{2}
  \sum_q
\beta_j^q
      a_{n+1}^q ,~~~{\ } n=2,4...,   \\
\int_0^1 dx x^n g_5^j(x,Q^2) &=&
\frac{1}{4}
  \sum_q
\beta_j^q
      a_{n}^q ,~~~{\ } n=1,3...~~.
\label{eqg1M}
\ea
Here $a_n^q$ is related to  the expectation value
$\langle pS|\Theta_{S}^{\beta\left\{
\mu_1 ... \mu_n\right\}}|pS \rangle$ (see refs.~\cite{BK1,BK2} for
a detailed description).
The factors $\alpha_j^q$ and $\beta_j^q$ are given by
\ba
\left(
\alpha_{|\gamma|^2}^q, \alpha_{|\gamma Z|}^q, \alpha_{|Z|^2}^q \right)
&=& \left [
e_q^2, 2 e_q g_V^q, (g_V^q)^2 + (g_A^q)^2 \right ] \\
\left(
\beta_{|\gamma Z|}^q, \beta_{|Z|^2}^q \right)
&=& \left [
2 e_q g_V^q,
2 g_V^q g_A^q \right ].
\label{eqAlBe}
\ea
Analogous relations to (\ref{eqYMP}--\ref{eqg1M})
 are derived for the charged current structure
functions~( cf.~ref.~\cite{BK2}).
By analytic continuation of the moment index $n$ into the complex plane
one obtains the following relations
between
the twist--2 contributions to the structure functions $g_k^j|_{k=1}^{5}$:
\ba
g_2^i(x) &=& -g_1^i(x) + \int_x^1\frac{dy}{y}g_1^i(x),
\label{qq7}  \\
g_4^j(x) &=& 2xg_5^j(x),
\label{qq8} \\
g_3^j(x)&=&4x\int_x^1\frac{dy}{y}g_5^j(y),
\label{qq9}
\ea
where $i=\gamma,\gamma Z, Z, W $ and $j=\gamma Z, Z, W $.
Eqs.~(\ref{qq7}) and (\ref{qq8}) are the Wandzura--Wilczek~\cite{WW}
and Dicus~\cite{DIC} relations, and eq.~(\ref{qq9}) is a {\it new}
relation which was firstly
 derived in ref.~\cite{BK1}.

\section{Covariant Parton Model}
%
\noindent
In the covariant parton model the hadronic
tensor for deep inelastic scattering is given by
\be
W_{\mu\nu, ab}(q,p,S)=\sum_{\lambda, i} \int d^4k
f_{\lambda}^{q_i}(p,k,S)
w_{\mu\nu ,ab, \lambda}^{q_i}(k,q) \delta[(k+q)^2-m^2].
\label{eq1}
\ee
Here $w_{\mu\nu, ab, \lambda}^{q_i}(k,q)$
denotes the hadronic tensor at the
quark level~\cite{BK1},
$f_{\lambda}^{q_i}(p,k,S)$ describes
the quark and antiquark
distributions  of the hadron,
$\lambda$ is the quark helicity,  $k$ the virtuality
of the initial state parton,
\be
k = xp + \frac{k^2 + k_{\perp}^2 - x^2 M^2}{2 x \nu} (q + xp) + k_{\perp},
\label{eq5}
\ee
and $m$ is the quark mass.

In the Bjorken limit
$Q^2, \nu \rightarrow \infty$,
$x = const.$ one obtains the following representation for the polarized
neutral current
structure functions including light quark mass effects:
\ba
g_1^j(x,\rho) &=&
\frac{\pi M^2x}{8} \sum_q \alpha_q^j
\int_{x+\frac{\rho}{x}}^{1+\rho}dy
\left [ x(2x-y) + 2 \rho \right ]
\tilde{h}_{q}(y,\rho),
\label{eqAA}
\\
g_2^j(x,\rho) &=&
\frac{\pi M^2}{8} \sum_q \alpha_q^j
\int_{x+\frac{\rho}{x}}^{1+\rho}dy
\left [x (2y-3x) -  \rho \right ]
\tilde{h}_{q}(y,\rho) \nonumber\\
&-&
\frac{\pi m^2}{4} \sum_q \gamma_q^j
\int_{x+\frac{\rho}{x}}^{1+\rho}dy
\tilde{h}_{q}(y,\rho),
\label{eqAC}
\label{eqVIO}
\\
g_3^j(x,\rho) &=& \frac{\pi M^2 x^2}{2}   \sum_q  \beta_q^j
\int_{x+\frac{\rho}{x}}^{1+\rho}dy
(y-x) \tilde{h}_{q}(y, \rho),
\\
g_4^j(x,\rho) &=& 2x g_5(x),
\label{eqAD}
\\
g_5^j(x,\rho) &=&
\frac{\pi M^2}{8} \sum_q \beta_q^j
\int_{x+\frac{\rho}{x}}^{1+\rho}dy
\left [x (2x-y) + 2 \rho \right ] \tilde{h}_q(y,\rho),
\label{eqAB}
\ea
with $\rho = m^2/M^2$,
$\gamma_q^j = g_{A_a}^q g_{A_b}^q, j~\equiv~ab$,
$ y=x+k^2_\bot/(xM^2)$,
and
\be
\tilde{h}_{q}(y, \rho)
=  \int dk^2 \hat{f}_{q}(y,k^2, \rho).
\label{eq11}
\ee
The corresponding relations for
charged current scattering are given in~\cite{BK2}.
The expressions
for $g_1^{em}$ and $g_2^{em}$
have been obtained in \cite{a17,a18}
already.

In the limit $ \rho
 \rightarrow 0$ the three relations
eqs.~(\ref{qq7}--\ref{qq9}) between the different structure functions
being
derived previously using the operator
product expansion
are obtained choosing $g_1$ and $g_5$ as the two basis  structure
functions
in lowest order QCD.
On the basis of the above relations a series of sum rules can be
derived. 
For $\rho \neq 0$ a new contribution to $g_2$ emerges for
$g_{A_a} g_{A_b} \neq 0$ leading to a corresponding violation
of the Burkhardt--Cottingham sum rule~\cite{BC}. The first  moment
of the structure functions $g_3$ and $g_4$ is predicted to be equal
for all current combinations also for $\rho \neq 0$ and
\begin{equation}
\int_0^1 dx~x \left [ g_1^k(x) + 2  g_2^k(x) \right ] = 0
\end{equation}
holds for arbitrary quark masses in the case of charged current
interactions. Also the Dicus relation~\cite{DIC} remains valid.
A detailed discussion is contained in refs.~\cite{BK1,BK2} where also 
numerical results are presented.

\section{Conclusions}
%
\noindent
We have derived the twist--2 contributions to the polarized structure
functions in lowest order QCD including weak currents. The results
obtained using the operator product expansion and the covariant
parton model agree. In lowest order two out of five structure functions
are independent
for the respective current combinations
and the
remaining structure functions are related by {\it three}
 linear operators.
A new relation between the structure functions $g_3^j$ and $g_5^j$ was
derived. As a consequence the first moment of $g_3^j$ and $g_4^j$ are
predicted to be equal.

The light quark mass corrections to the structure functions
$\left.
g^j_k \right|_{k=1}^5$ were calculated in the covariant parton model.
The first moments of the structure functions $g_3$ and $g_4$ are
equal also in the  presence of the
quark mass corrections. The Dicus relation remains to be valid.
The Burkhardt--Cottingham sum rule is broken by a term
$\propto g_{A_a} g_{A_b} m^2/M^2$, i.e. for pure $Z$~exchange and in
charged current interactions.


\end{document}